\documentclass[aps,prl,reprint,superscriptaddress]{revtex4-2}

\usepackage{graphicx}
\usepackage{dcolumn}
\usepackage{bm}

\usepackage{siunitx}
\usepackage{amsmath}
\usepackage{color}
\usepackage{ulem}

\begin{document}

\title{Free-Running Long-Distance Reference-Frame-Independent Quantum Key Distribution}

\author{Bang-Ying Tang}
\thanks{these authors contributed equally to this work}
\affiliation{College of Computer Science and Technology, National University of Defense Technology, Changsha 410073, China}

\author{Huan Chen}
\thanks{these authors contributed equally to this work}
\affiliation{College of Liberal Arts and Sciences, National University of Defense Technology, Changsha 410073, China}

\author{Ji-Peng Wang}
\affiliation{School of Science and State Key Laboratory of Information Photonics and Optical Communications, Beijing University of Posts and Telecommunications, Beijing 100876, China}

\author{Hui-Cun Yu}
\affiliation{Information and Navigation College, Air Force Engineering University, Xi’an 710077, China}
\affiliation{College of Advanced Interdisciplinary Studies, National University of Defense Technology, Changsha 410073, China}

\author{Lei Shi}
\affiliation{Information and Navigation College, Air Force Engineering University, Xi’an 710077, China}

\author{Shi-Hai Sun}
\affiliation{School of Physics and Astronomy, Sun Yat-sen University, Zhuhai 519082, China}

\author{Wei Peng}
\affiliation{College of Computer Science and Technology, National University of Defense Technology, Changsha 410073, China}

\author{Bo Liu}
\email[]{liubo08@nudt.edu.cn}
\affiliation{College of Advanced Interdisciplinary Studies, National University of Defense Technology, Changsha 410073, China}

\author{Wan-Rong Yu}
\email[]{wlyu@nudt.edu.cn}
\affiliation{College of Computer Science and Technology, National University of Defense Technology, Changsha 410073, China}

\begin{abstract}

Rapidly and randomly drifted reference frames will shorten the link distance and decrease the secure key rate of realistic quantum key distribution (QKD) systems. However, an actively or inappropriately implemented calibration scheme will increase complexity of the systems and may open security loopholes. In this article, we present a free-running reference-frame-independent (RFI) QKD scheme, where measurement events are classified into multiple slices with the same misalignment variation of reference frames and each slice performs the post-processing procedure individually. We perform the free-running RFI QKD experiment with a fiber link of \SI{100}{km} and the misalignment of the reference frame between Alice and Bob is varied more than 29 periods in a 50.7-hour experiment test. The average secure key rate is about \SI{734}{bps} with a total loss of \SI{31.5}{dB}, which achieves the state-of-art performance of the long-distance RFI QKD implementations. Our free-running RFI scheme can be efficiently adapted into the satellite-to-ground and drone based mobile communication scenarios, as it can be performed with rapidly varying reference frame and a loss more than \SI{40}{dB}, where no secure key can be obtained by the original RFI scheme.

\end{abstract}
\maketitle
Quantum key distribution (QKD) has the potential to generate the information-theoretical-secure keys between distant users (Alice and Bob)~\cite{RN4, RN445, RN446}. Since the first BB84 protocol proposed~\cite{RN450}, QKD has stepped into realistic applications from the laboratory~\cite{RN636, RN563, RN584, RN645, RN553, RN562, RN568}. The communication distance of QKD over different communication scenarios is significantly improved, such as \SI{509}{km} with fiber link~\cite{RN684}, \SI{4600}{km} space-to-ground communication network~\cite{RN1971}, \SI{1}{km} optical-relayed entanglement distribution over drones~\cite{RN681}.

In general, the realistic QKD systems have a common need of establishing a shared reference frame between distant users by characterizing and calibrating the polarization or phase of photons. Nevertheless, an actively or inappropriately implemented calibration scheme may increase the complexity of the systems and may open an unpredictable security loophole~\cite{RN2021, RN678}.

Reference-frame-independent (RFI) QKD protocol, proposed in 2010, is robust for the slowly varying reference frames and suitable to the satellite-to-ground and drone-based systems~\cite{RN569}. The secure key rate of RFI QKD systems varies a lot with given misalignment ($\theta _0$) of reference frames~\cite{RN654}. Furthermore, F. Wang \textit{et al.} shows the valid condition of RFI-QKD is that the misalignment variation $\Delta \theta \leq \pi$~\cite{RN655}. With the assumption of $\Delta \theta \approx 0$, the fiber link distance of RFI QKD reaches to \SI{160}{km} implemented with measurement-device-independent scheme~\cite{RN568, RN637} and the free space RFI QKD experiment was demonstrated with total loss of \SI{38.0}{dB}~\cite{RN553, RN661}. However, the misalignment of reference frames in realistic RFI QKD scenarios varies randomly, and the communication distance will be reduced heavily. W. Liang \textit{et al.} performed a proof-of-principle experiment of RFI QKD with random $\theta_0$ and slowly varying phase environment, nevertheless, the communication distance was limited to \SI{65}{km} when considering the finite-length effects~\cite{RN663, RN675}. 

In this article, we proposed a free-running RFI scheme that can be adapted to the long-distance QKD systems with randomly and rapidly varying misalignment of reference frames. We implemented a free-running RFI QKD experiment over \SI{100}{km} fiber link with a total loss of \SI{31.5}{dB} and the system repetition rate of \SI{80}{MHz}. In the experiment, time-bin encoded photons in $Z$ basis are used for generating secure keys and the phase-encoded photons in the $X$ ($Y$) basis are used for quantifying the characteristics of the quantum link. The experiment ran for \SI{50.7}{hours} and the misalignment of reference frame between Alice and Bob was more than 29 periods. The secure key rate of our implementation reaches \SI{734}{bps}, which achieves the state-of-art performance of the long-distance RFI QKD systems.

\begin{figure}[htb]
	\centering
	\includegraphics[width=0.5\textwidth]{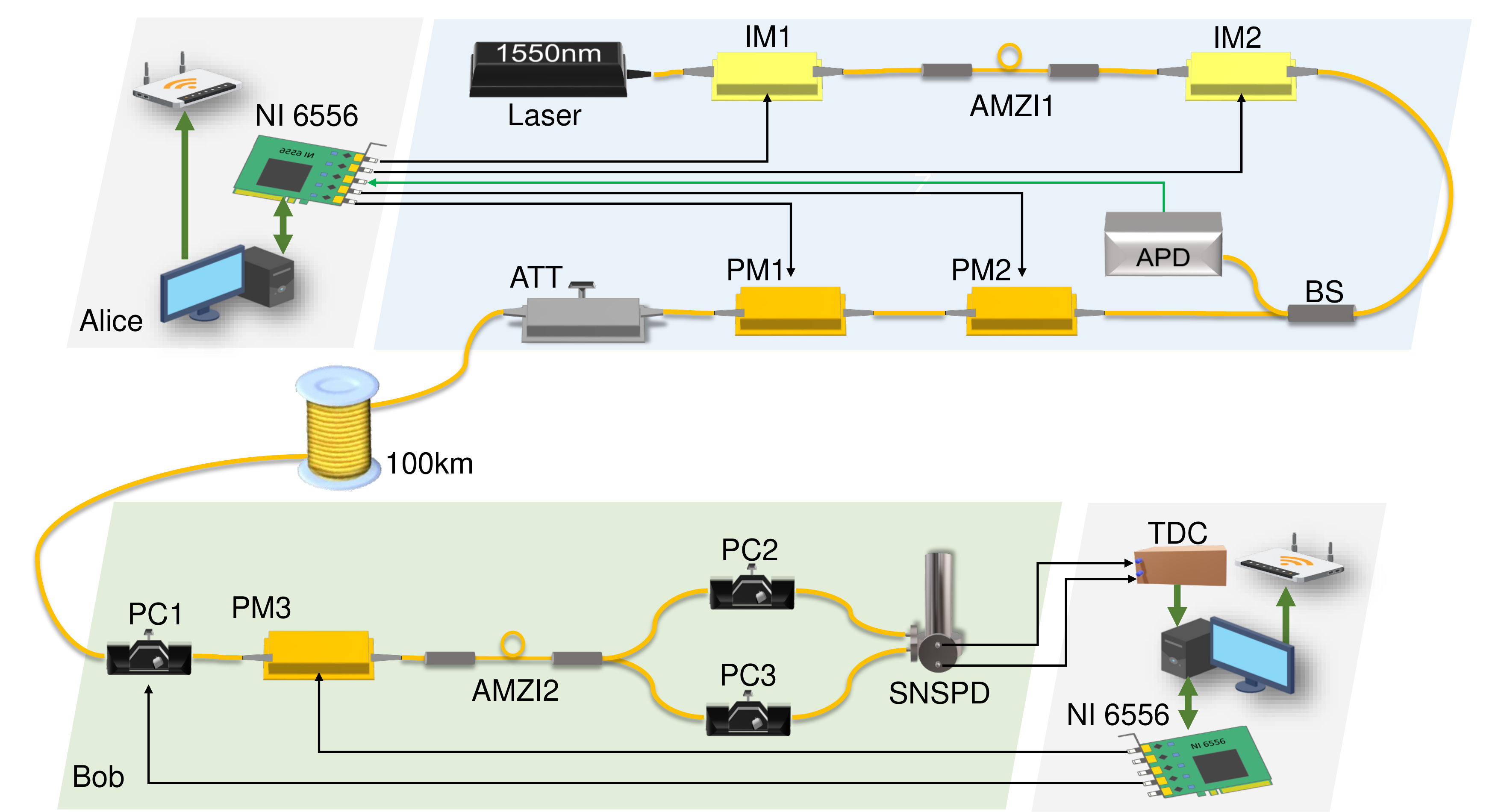}
	\caption{\label{fig:setup_exp} Illustration of the \SI{100}{km} free-running RFI QKD experiment. Two NI 6556 boards control the lasers and modulators. The computers are used for data acquisition, post-processing, and classical communication. APD is used to measure the intensity of signal and decoy states so that NI 6556 adjusts the control signal of IM1. IM: intensity modulator, BS: beam splitter, PM: phase modulator, PC: polarization controller, AMZI: asymmetric Mach-Zehnder interferometer, SNSPD: superconducting nanowire single photon detector, TDC: Time-to-Digital converter, APD: avalanche photon diode, NI 6556: National Instruments 6556.}
\end{figure}

The setup of our free-running RFI QKD experiment is shown in FIG.~\ref{fig:setup_exp}. At Alice's side, photons are generated by a homemade laser with a repetition rate of \SI{80}{MHz} and a pulse width of \SI{1.0}{ns}. An intensity modulator (IM1) and an attenuator (ATT) are used to modulate quantum (signal and decoy) states of the corresponding intensity. An asymmetric Mach–Zehnder interferometer (AMZI1) divides each pulse into two adjacent pulses ($s$ and $l$) with a separation time of \SI{4.6}{ns}. The intensity modulator IM2 prepares the bit $0$ or $1$ of $Z$ basis by suppressing $s$ or $l$ randomly. When both $s$ and $l$ pulses pass through IM2, a phase of $\{ 0, \pi /2, \pi, 3\pi /2 \}$ will be randomly added to $s$ pulse with two phase modulators (PM1 and PM2), where $0$ and $\pi$ indicate bit $0$ and $1$ in $X$ basis, $\pi / 2 $ and $3 \pi / 2$ indicate bit $0$ and $1$ in $Y$ basis. Then, pulses are sent to Bob through a \SI{100}{km} optical fiber with a loss of \SI{0.235}{dB/km}.

At Bob's side, the detection basis ($X$ and $Y$) is randomly chosen by the phase modulator (PM3), where a phase of $0$ ($\pi /2$) is added to the $s$ pulse. Then another asymmetric Mach–Zehnder interferometer (AMZI2) which has the same path difference with the AMZI1 divides each pulse into two pulses, resulting four pulses, \{$s + s$, $s + l$, $l + s$, $l + l$\}. $s+l$ and $l+s$ pulses will merge into one pulse due to the same arriving time and interfere at AMZI2. The interference determines the exiting port of the $s+l$ ($l+s$) pulse at the AMZI2, indicating the bit $0$ or $1$ of $X$($Y$) basis, respectively. $s + s$ and $l + l$ pulses represent bit $0$ and $1$ of $Z$ basis. Then, photons are coupled to two channels of the superconducting nanowire single photon detector (SNSPD), and the arriving time is recorded with a time-to-digital converter (TDC). The optimal detection efficiency, about 80\%, is ensured by the polarization controllers (PC2 and PC3). The dark count rate of the SNSPD is about $200$ per second. The programmable modules (National Instruments 6556) are used to control the laser and modulators. The total loss of Bob's receiving module is around \SI{10}{dB}.

Furthermore, an active feedback mechanism is performed to stabilize the intensity of the quantum state, where a beam splitter (BS) and an avalanche photon detector (APD) are performed to monitor the intensity of quantum states. The optimal detection parameters are analyzed from the recorded events of Bob's side and then feedback to the programmable polarization controller (PC1) and PM3.

In the RFI QKD system, the reference frame of $X$ ($Y$) is defined as 
\begin{equation}
	\begin{aligned}
		X_B &=X_A \cos \theta + Y_A \sin \theta,\\
		Y_B &=Y_A \cos \theta - X_A \sin \theta,
	\end{aligned}
\end{equation} 
where $\theta$ is the misalignment angle of reference frames between Alice and Bob and $\theta \in [0,2\pi)$. The quality of the quantum channel  $C$ can be estimated as~\cite{RN569}
\begin{equation}
	\begin{aligned}
		C &= \left \langle X_A X_B \right \rangle ^2 + \left \langle Y_A Y_B \right \rangle ^2  + \left \langle X_A Y_B \right \rangle ^2  + \left \langle Y_A X_B \right \rangle ^2.
	\end{aligned}
\end{equation}

Given initial $\theta_0$, the RFI-QKD can be performed with the varying $\Delta \theta$ which satisfies
\begin{equation}
	\frac{3\times 10^4}{\eta N_0} \omega \leq \Delta \theta \leq \pi,
	\label{equ:rangeOfTheta}
\end{equation}
where $\eta$ is the total efficiency of system, $\omega$ denotes the intense constant of noise and $N_0$ is the photon-generating rate~\cite{RN655}.

Assume $E_k^{\alpha,\beta}$ is the QBER, where $k \in \{\mu, \nu\}$ is the mean photon number of signal and decoy states, $\alpha, \beta \in \{X, Y, Z\}$ are the basis chosen by Alice and Bob, respectively. And the total pulses sent by Alice is $N_\textrm{t}$.

In the free-running RFI scheme, we split the misalignment angle $\theta$ into $m$ slices as 
\begin{equation}
	\Theta_i=\left\{ \begin{matrix}
 \{\theta | (2-\frac{1}{m})\pi \leq \theta < 2\pi~ or ~ 0\leq \theta < \frac{\pi}{m}\}, & i=0 \\ 
 \{\theta | \frac{2i-1}{m}\pi \leq \theta < \frac{2i+1}{m}\pi\}, & i>0
 \end{matrix}\right. ,
\end{equation}
where $i=0,1,\dots,m-1$. 

According to Equ (\ref{equ:rangeOfTheta}), the interval sampling time $T$ for quantifying the quantum channel should be less than $\pi /\omega$.

Given $T$, $\theta$ of each interval can be calculated as 
\begin{equation}
	\theta=\left\{\begin{array}{cc}
		\arccos \left(1-2 E_{\mu}^{X X}\right) & E_{\mu}^{X Y} < 0.5 \\
		2 \pi-\arccos \left(1-2 E_{\mu}^{X X}\right) & E_{\mu}^{X Y} \geq 0.5
		\end{array}\right. .
\end{equation}

Thus, the measurement events can be split into $m$ slices according to the measured QBER in XY-basis and then generate the secure keys for each slice respectively.

\begin{figure*}[htb]
	\centering
	\includegraphics[width=0.85\textwidth]{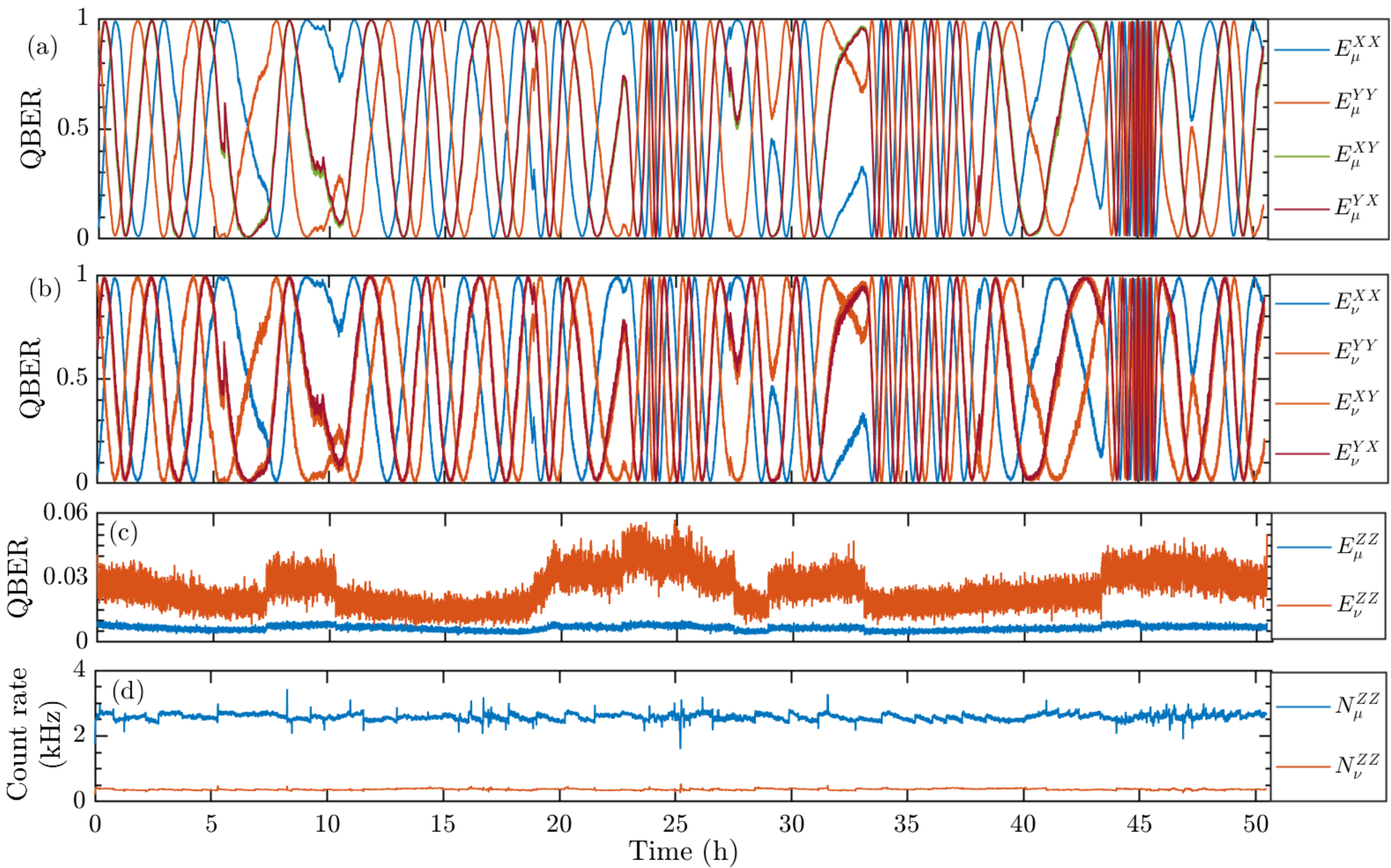}
	\caption{\label{fig:error_rate_raw}The measured QBER in each bases, and the count rate in Z-basis. Each point is measured within 5s. (a) [(b)] shows the QBER of signal [decoy] state on $X$, $Y$ basis for the total \SI{50.7}{hours}, (c) shows the QBER of signal state and decoy state on $Z$ basis: $E_\mu^{ZZ}$ and $E_\mu^{ZZ}$, and (d) shows the count rate of signal and decoy state on $Z$ basis: $N_\mu ^{ZZ}$ and $N_\nu ^{ZZ}$.} 
\end{figure*}

\begin{figure}[htb]
	\centering
	\includegraphics[width=0.48\textwidth]{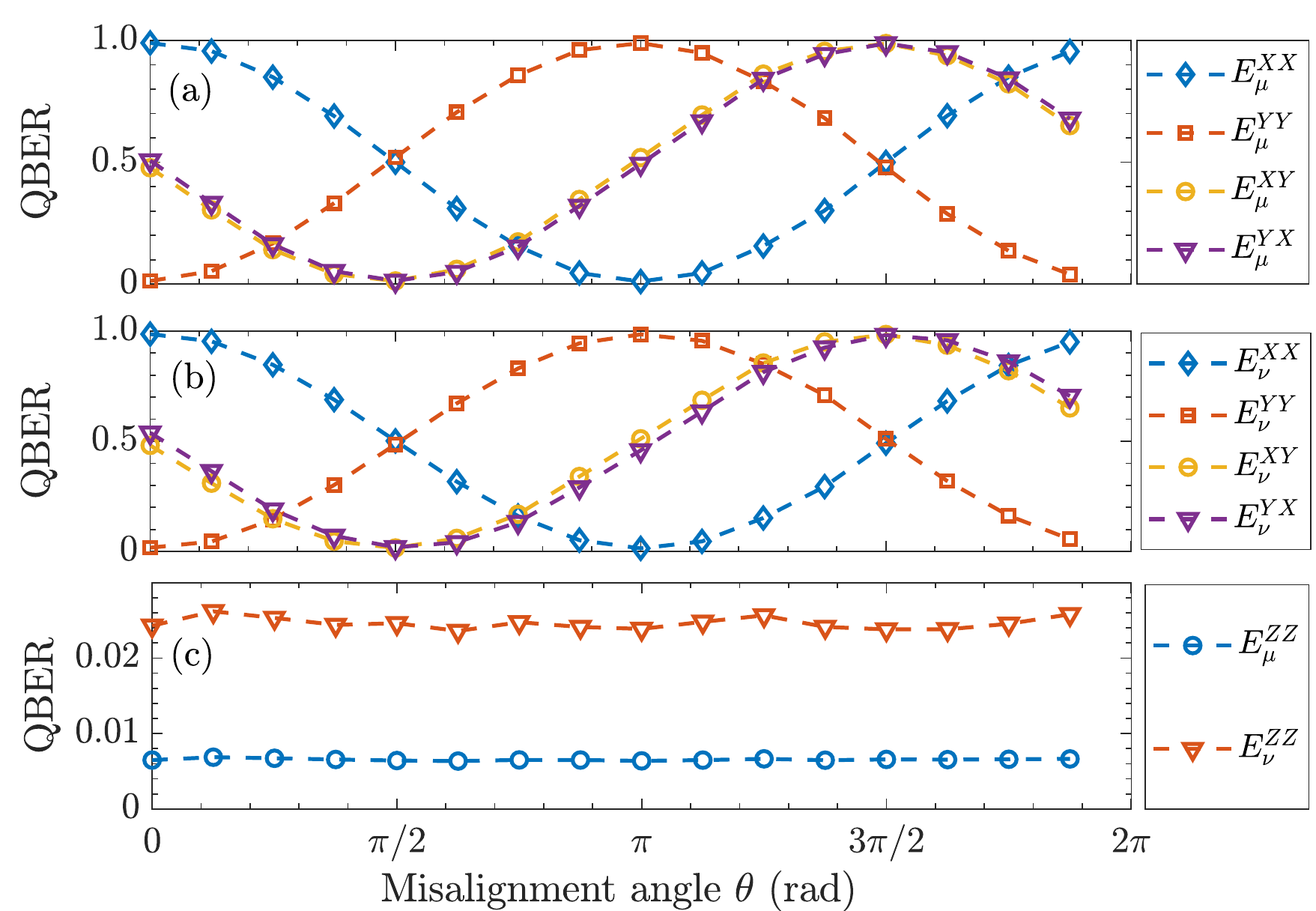}
	\caption{\label{fig:error_rate_sub}The QBER of the merged sets with the experimental measured data. (a) shows the QBER of signal state on $X$,$Y$ basis, (b) shows the QBER of decoy state on $X$,$Y$ basis, (c) shows the QBER on $Z$ basis, and each point represents the merged result of each set.}
\end{figure}

\begin{figure}[htb]
	\centering
	\includegraphics[width=0.48\textwidth]{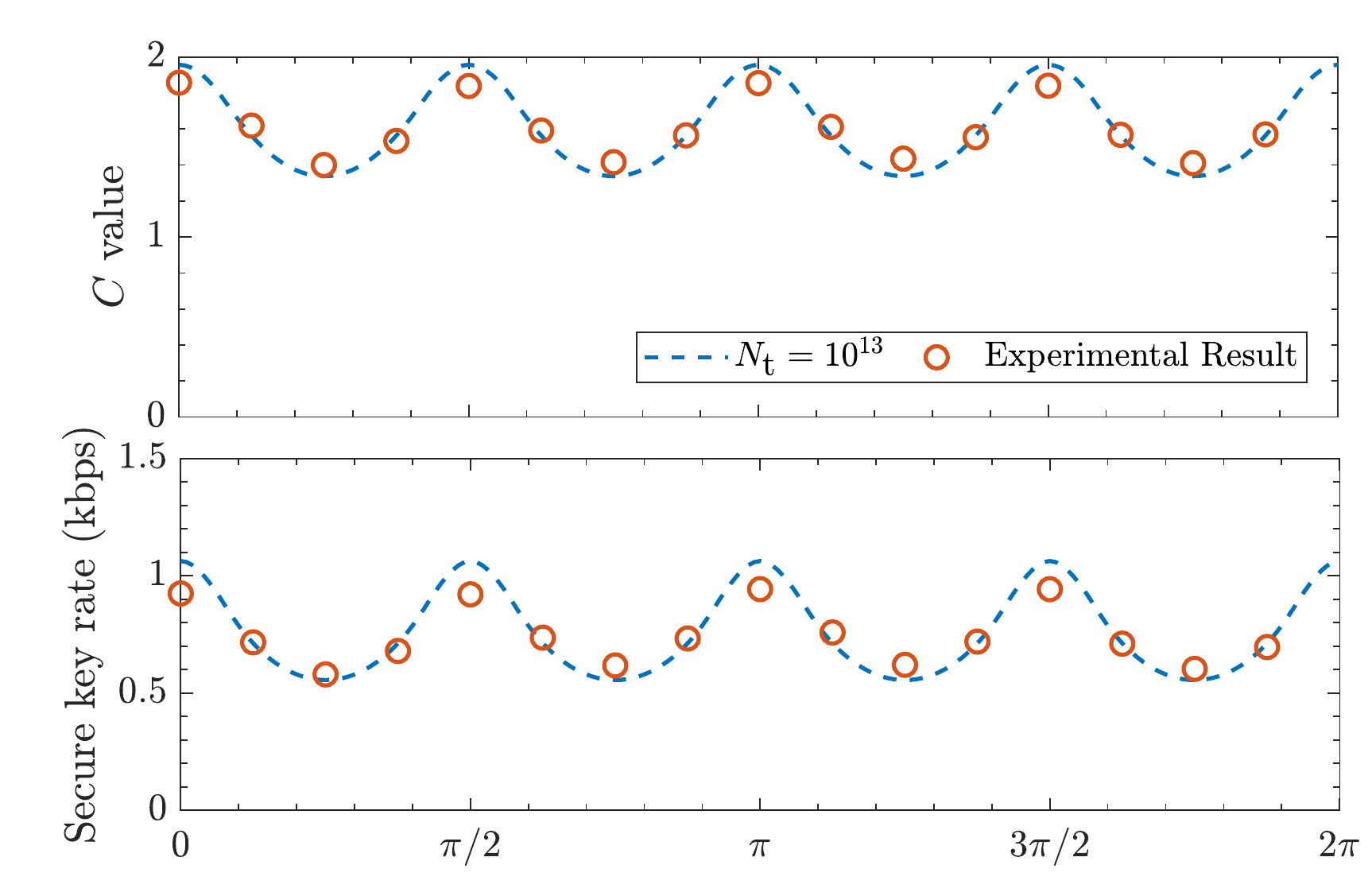}
	\caption{\label{fig:skr_sub}  The $C$ value, secure key rate of merged sub-block and the simulated result with $N_\textrm{t} = 10^{12}$ and $\Delta \theta = \pi / 8$. The $C$ value and secure key rate varies with $\theta$ because the misalignment of reference frame affects the leaked information~\cite{RN654}.}
\end{figure}

 In the experiment, the $Z$ basis is sent with the probability of 50\%, $X$ and $Y$ basis are sent with the probability of 25\%. The signal, decoy, and vacuum states are sent with the probability of 50\%, 40\%, and 10\%, respectively. The mean photon number of signal and decoy states are $0.722$ and $0.104$. The experiment is free-running for about \SI{50.7}{hours}. The sampling interval $T$ is set to 5 seconds. The slice number $m$ is set as $16$ and the error correction efficiency is $1.16$~\cite{RN490}.

\begin{figure}[htb]
	\centering
	\includegraphics[width=0.48\textwidth]{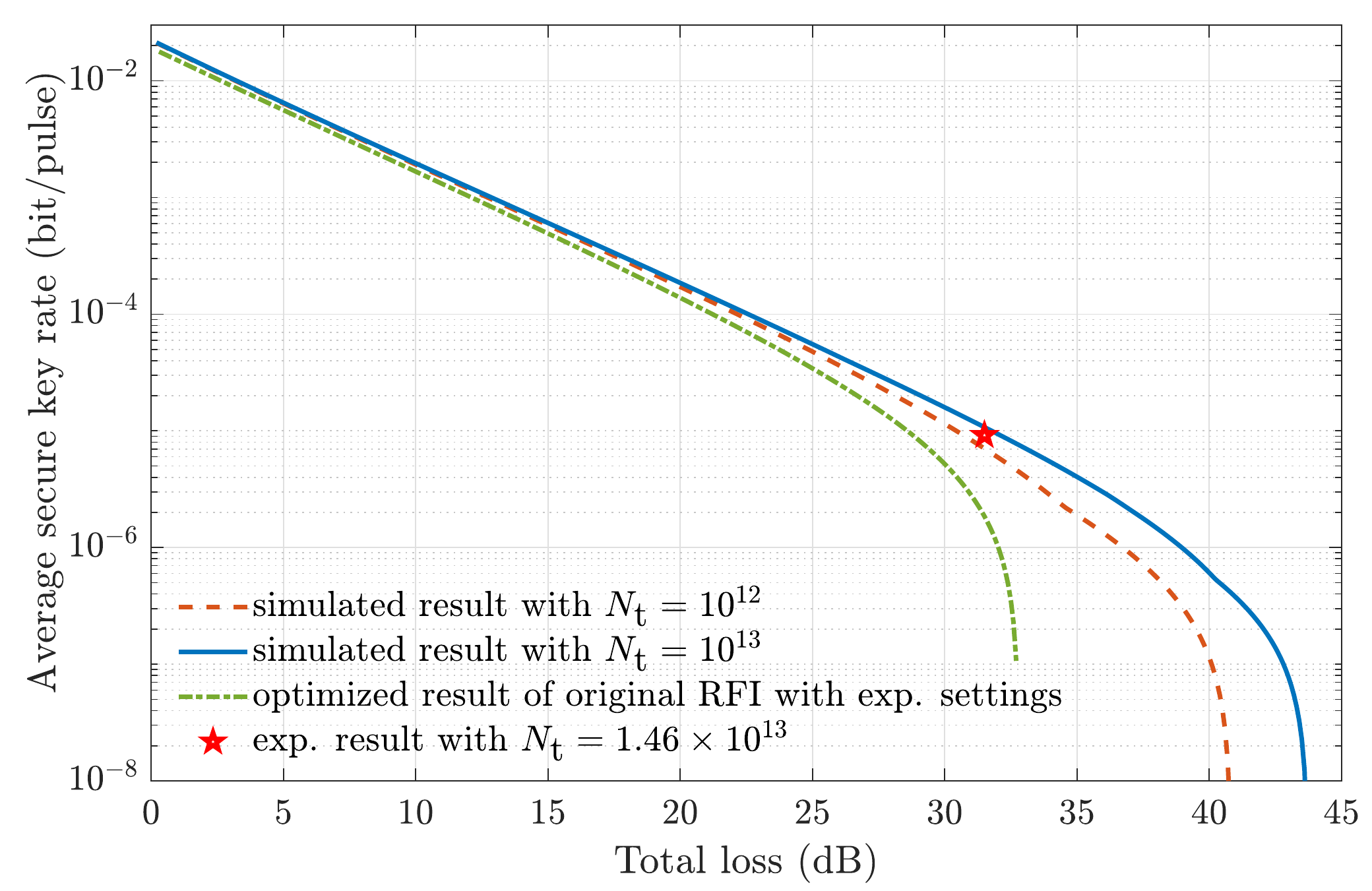}
	\caption{\label{fig:finite_len_skr}The simulated secure key rate of our free-running RFI scheme with $N_\textrm{t}=10^{12}, 10^{13}$. Our experimental result and the simulated optimal secure key rate of the original RFI scheme are also shown here. In the simulation, $\omega = 6.9 \times 10^{-3}$~rad/s and $m=16$. }
\end{figure}

FIG.~\ref{fig:error_rate_raw} shows the QBER of signal and decoy states for each bases and the count rate of $Z$ basis. The QBER of the X, Y bases varied periodically with the time due to the environmental noise which affects the phase of the two interferometers, while the QBER of the Z basis maintained stable with small fluctuations. The average QBER of the signal state on the $Z$ basis is about 0.6\%. The average count rate of the signal and decoy states of $Z$ basis is about \SI{2.6}{kbps} and \SI{0.4}{kbps}. During the whole experiment, the reference frame varied for more than 29 periods and the noise $\omega$ reaches an upper bound of $6.9 \times 10^{-3}$~rad/s.

Without loss of generality, we use $\{0, 2\pi/m,\cdots,2(m-1)\pi/m\}$ to represent the misalignment angle of the slices. FIG.~\ref{fig:error_rate_sub} shows the average QBER of each slice. The QBERs of $X$, $Y$ basis vary against the rotation $\theta$ while the QBERs of signal and decoy state on $Z$ basis are stabilized at around 0.6\% and 2.3\%. The corresponding $C$ value and secure key rate of each slice are shown in FIG.~\ref{fig:skr_sub} with red circles. The blue dash lines in FIG.~\ref{fig:skr_sub} illustrate the simulated results with $N_\textrm{t}=10^{13}$~\cite{RN562,RN636,RN565,RN654}. In principle,  the $C$ value and the secure key rate reach the maximum value with $\theta \in \{0,\pi/2, \pi, 3\pi/2 \}$ and drop to the minimum value with $\theta \in \{\pi /4,3\pi /4,5\pi /4,7\pi /4\}$. The experimental $C$ value and secure key rate are slightly less than the simulated results due to the imperfect preparation and measurement of photons. The secure key rate is varying between \SI{603}{bps} and \SI{943}{bps}, in slice $[15\pi/16,17\pi/16)$ and $[3\pi/16,5\pi/16)$ respectively. Finally, the measured average secure key rate in our experiment is \SI{734}{bps}.

Furthermore, we simulate the average secure key rate of the free-running RFI QKD versus the total loss with $N_{\textrm{t}}=10^{12}, 10^{13}$ and the noise $\omega = 6.9\times 10^{-3}$~rad/s. The results are shown in FIG.~\ref{fig:finite_len_skr}. The total loss tolerance of the original RFI QKD protocol is limited to \SI{32.6}{dB}, which can be increased to around \SI{43.3}{dB} with our free-running RFI scheme. Meanwhile, the free-running RFI scheme is conducted after the basis shifting procedure, and the number $m$ of slices can be optimized to further improve the performance. Especially, our free-running RFI scheme can be performed into the satellite-to-ground and drone-based QKD scenarios, which avoids the complicated alignment of unpredictable varying reference frames. 

This work was supported by National Natural Science Foundation of China under Grant No. 61972410 and No. 61971436, the Research Plan of National University of Defense Technology under Grant No. ZK19-13 and No. 19-QNCXJ-107 and the Postgraduate Scientific Research Innovation Project of Hunan Province under Grant No.CX20200003. 


%

\end{document}